\documentclass[conference]{IEEEtran}

\usepackage{cite}
\usepackage{amssymb}

\ifCLASSINFOpdf
   \usepackage[pdftex]{graphicx}
\else
   \usepackage[dvips]{graphicx}
\fi

\usepackage[cmex10]{amsmath}
\DeclareMathOperator{\He}{H}
\DeclareMathOperator{\Tr}{T}
\newcommand{\B}[1]{\mathbf{#1}}

\IEEEoverridecommandlockouts

\begin{document}

\title{Adaptive Beamforming in Interference Networks via Bi-Directional Training}

\author{\IEEEauthorblockN{Changxin Shi, Randall A. Berry, and Michael L. Honig }
\IEEEauthorblockA{Department of Electrical Engineering and Computer Science\\
Northwestern University, Evanston, Illinois 60208\\
Email: cshi@u.northwestern.edu, \{rberry, mh\}@eecs.northwestern.edu}
\thanks{This research was supported in part by ARO under grant W911NF-06-1-0339 and DARPA under grant W911NF-07-1-0028. This is a revised version of a paper that appeared in 44th Annual Conference on Information Sciences and Systems (CISS), Princeton, NJ, March 2010. The major change in this version is that we consider an additional receive filter update during the data transmission to further improve the performance.}
}


\maketitle

\begin{abstract}
We study distributed algorithms for adjusting beamforming vectors and
receiver filters in multiple-input multiple-output (MIMO) interference
networks, with the
assumption that each user uses a single beam and a
linear filter at the receiver. In such a setting there
have been several distributed algorithms studied for maximizing the
sum-rate or sum-utility assuming perfect channel state information
(CSI) at the transmitters and receivers. The focus of this paper is
to study adaptive algorithms for
time-varying channels, without assuming any CSI at the transmitters or
receivers. Specifically, we consider an adaptive version of the recent
Max-SINR algorithm for a time-division duplex system. This algorithm
uses a period of bi-directional training followed by a block of data transmission. Training in the forward direction is sent
using the current beam-formers and used to adapt the receive
filters. Training in the reverse direction is sent using the current
receive filters as beams and used to adapt the transmit beamformers.
The adaptation of both receive filters and beamformers is done using a
least-squares objective for the current block. In order to improve
the performance when the training data
is limited, we also consider using exponentially weighted data from
previous blocks. Numerical results are presented that compare
the performance of the algorithms in different settings.
\end{abstract}

\section{Introduction}\label{introduction}

The use of multiple antennas in wireless networks offers the promise
of mitigating interference and enabling high spectral efficiencies.
However achieving these benefits in a decentralized network
requires distributed algorithms for coordinating the pre-coding matrices
used by each transmitter with minimal overhead. A number of such
algorithms have been studied for multiple-input multiple-output (MIMO)
interference networks including those in
\cite{ICC09,Asilomar09,JafarGlobecom08} under the assumption
that transmitters have perfect channel state information (CSI)
for channel matrices to all receivers. Our focus in this work is to
relax this assumption and develop adaptive
algorithms for time-varying channels without assuming any CSI
at the transmitters or receivers.

We consider a MIMO interference network with block fading.
Each transmitter uses
a rank one pre-coding matrix, i.e. a beamformer, and the
receivers are assumed to be linear, with all interference
treated as noise, so that
the rate is determined by the received signal-to-interference plus
noise ratio (SINR). Our objective is to design an adaptive distributed
algorithm for  updating the precoders and beamformers to maximize the
sum-rate while accounting for the needed overhead. To accomplish this,
we consider an algorithm for a synchronous
time-division duplex (TDD) system that uses one or more
bi-directional training
periods  at the beginning of
each block. Each period consists of a forward phase followed by a
reverse phase. During the forward phase, all
transmitters simultaneously send pilots using their current
beamformers and each receiver updates their receive filters.
During the backward phase, the receivers
transmit pilots, using the current receive filter as a beamformer and
the transmitters update their beamformers. The updates during each
phase are used to directly adapt the beamformers and receive filters
based on a standard least-squares criterion.
For a given number of channel users per coherence block, we consider
the net throughput achieved after subtracting the channel uses
used for training and via numerical results study the effect
of varying the training length on these metric. For short coherence
blocks, the optimal training length becomes small leading to
imprecise estimation and degraded performance. For such settings, we
also study a recursive variation of our algorithm based on using
exponentially weighted data from previous blocks.

The algorithms we study are inspired by the Max-SINR algorithm in
\cite{JafarGlobecom08}. This algorithm also iterates between
transmitter and receiver updates assuming perfect CSI
and static channels. Though no convergence proof for this algorithm
is given in \cite{JafarGlobecom08}, numerical results show that it
has good performance and at high SNRs can achieve the optimal
multiplexing gain by successfully aligning the interference at
each receiver~\cite{CadambeJafarIT08}. Related iterative algorithms
(also assuming perfect CSI) are given in
\cite{PetersHeath09,Asilomar09}. The bi-directional training method
presented here is also related to schemes for
two-way channel estimation presented in
\cite{Mar06,GomPapSun08,OsaMurYamYos09,SteSab08,WitTayWar08,ZhoLamSad10}.
The key difference here is that we are directly estimating the optimal
beamformer and receive filter as opposed to estimating the CSI needed
to compute those coefficients.

\section{System Model}\label{system model}
We consider a peer-to-peer wireless network with $K$ transmitter-receiver
pairs (henceforth referred to as {\it users}) communicating
through MIMO links, sharing the same spectrum.
Each transmitter has $N_T$ antennas, each receiver has $N_R$ antennas,
and the channel from the $k$-th transmitter to the $j$-th receiver
is denoted by a complex matrix $\B{H}_{jk}\in \mathbb{C}^{N_R\times N_T}$.
The channels are assumed to be block-fading, i.e.,
$\B{H}_{jk}$ remains constant for
$L$ symbol periods and jumps to another value for
the next $L$ symbols, according to the update
\begin{equation}
\B{H}_{jk}^{(n)} = \alpha\B{H}_{jk}^{(n-1)}+\sqrt{1-\alpha^2}\B{W}_{jk}^{(n)} \label{eq:blockfading}
\end{equation}
where $n$ is the block index and $\B{W}_{jk}^{(n)}$ is a matrix with
{\it i.i.d} complex Gaussian entries having zero mean
and the same variance as $\B{H}_{jk}$. (The index $n$ will sometimes be
omitted if no confusion results.)
We assume that neither
transmitters nor receivers have {\em a priori} channel information.

For simplicity, we assume each transmitter transmits only one beam
to its desired receiver, i.e., the precoding matrix has rank one.
The beamforming vector for transmitter $k$ is
$\B{v}_k\in \mathbb{C}^{N_T}$ and satisfies
the power constraint $\lVert\B{v}_k\rVert^2 \leq P_k^{max}$.
The $i$-th received signal vector at the $k$-th receiver in
the $n$-th block is then given by
\begin{equation}
\B{y}_k (i) = \B{H}_{kk}^{(n)}\B{v}_kx_k (i) +
\sum_{j\neq k}\B{H}_{kj}^{(n)}\B{v}_jx_j (i) + \B{n}_k (i)
\label{eq:receivedsignal}
\end{equation}
where $x_k$ is the unit variance data symbol from
transmitter $k$ and $\B{n}_k$ is the additive noise with
covariance matrix $E[\B{n}_k\B{n}_k^{\He}] = \B{R}_k$.

We assume linear receivers, so that the estimated symbol for user $k$ is
\begin{equation}
 \widehat{x}_k = \B{g}_k^{\He}\B{y}_k\label{eq:estimation}
\end{equation}
where $\B{g}_k$ is the corresponding receive filter.
Given a set of beamformers $\B{v}_k$'s and receive
filters $\B{g}_k$'s, the
Signal-to-Interference-plus-Noise Ratio (SINR) for user $k$
can be written as
\begin{equation}
\gamma_k = \frac{\lvert\B{g}_k^{\He}\B{H}_{kk}^{(n)}\B{v}_k\rvert^2}{\sum_{j\neq k}\lvert\B{g}_k^{\He}\B{H}_{kj}^{(n)}\B{v}_j\rvert^2 + \lvert\B{g}_k^{\He}\B{R}_k\B{g}_k\rvert}.
\end{equation}

Ideally, we would like to choose the set of beamforming vectors
$\{\B{v}_k\}$ and set of receive filters $\{\B{g}_k\}$
to maximize the sum rate $\sum_{k=1}^K \log (1+\gamma_k )$
within each block. This is complicated by the assumption
that the channels are time-varying, and initially unknown
at all nodes. Estimating the channels requires overhead, which
reduces the rate, and furthermore, if the channels vary too
quickly, then the channel estimates are likely to be inaccurate.
Therefore we desire an estimation scheme with minimal overhead,
and which adapts to the time-variations of the channel.

\section{Bi-Directional Optimization: Max-SINR Algorithm}
The adaptive bi-directional training scheme to be described is based
on the Max-SINR algorithm presented in \cite{JafarGlobecom08}.
That algorithm iteratively optimizes the transmit precoders
and receivers, assuming the transmitters/receivers
each know their direct- and cross-channel matrices.
It consists of the following steps:
(i) Fix the precoders and optimize the receivers;
(ii) Reverse the direction of transmission, so that the roles
of the receiver filters and precoders are swapped, and
optimize the precoders (now the receivers).

The optimization criterion in each step is the associated SINR, i.e.,
in step (i) the receiver for user $k$ is obtained by solving
\begin{align}
\label{gk}
 \max_{\B{g}_k}&\quad\frac{\lvert\B{g}_k^{\He}\B{H}_{kk}\B{v}_k\rvert^2}{\sum_{j\neq k}\lvert\B{g}_k^{\He}\B{H}_{kj}\B{v}_j\rvert^2 + \lvert\B{g}_k^{\He}\B{R}_k\B{g}_k\rvert}\\
 \mathrm{s.t.}&\qquad\qquad\lVert\B{g}_k\rVert^2 = P_k^{max}
\end{align}
and in step (ii) the beamformer for user $k$ is updated by solving
\begin{align}
\label{vk}
 \max_{\B{v}_k}&\quad\frac{\lvert\B{v}_k^{\He}\B{H}_{kk}^{\He}\B{g}_k\rvert^2}{\sum_{j\neq k}\lvert\B{v}_k^{\He}\B{H}_{jk}^{\He}\B{g}_j\rvert^2 + \lvert\B{v}_k^{\He}\B{R}_k\B{v}_k\rvert}\\
 \mathrm{s.t.}&\qquad\qquad\lVert\B{v}_k\rVert^2 = P_k^{max}.
\end{align}

Although inspired by a duality type of argument, which applies to
the uplink/downlink, the max-SINR method does not appear to maximize
a particular objective. Hence so far, there is no proof that the
algorithm converges. Nevertheless, numerical results show
that for the scenario considered (i.e., one beam per user)
the Max-SINR algorithm essentially achieves the maximum sum rate
over a wide range of SNRs \cite{Asilomar09}.


\section{Bi-Directional Training}
Maximizing the received SINR in \eqref{gk} and \eqref{vk}
is equivalent to minimizing the Mean Squared Error (MSE)
at the output of the corresponding filter.
This leads to an adaptive version in which the MSE is
replaced by a Least Squares (LS) cost function.
Here we assume that in each step the set of transmitters
or the set of receivers synchronously transmit training
sequences in each direction.


Specifically, in the $n$-th block, we assume that
the transmitters synchronously transmit
the sequence of $M$ training symbols given by
the matrix $\B{B}^{\He}$ where
$\B{B} = [\B{b}_1^{\He} ,\cdots,\B{b}_K^{\He}]$
and $\B{b}_k$ is the $1\times M$ row vector containing
the training symbols $b_k (1),\cdots,b_k(M)$.
The received signal at receiver $k$ is then given by \eqref{eq:receivedsignal}
where $x_k (i) = b_k (i)$. At receiver $k$
the estimated symbol at time $i$ is then
$\hat{b}_k (i) = \B{g}_k^{\He} \B{y}_k (i)$.
The corresponding sequence of estimated symbols is
$\widehat{\B{b}}_k = \B{g}_k^{\He}\B{Y}_k$, where
\begin{equation}
\B{Y}_k = \Big[\B{y}_k (1),\cdots,\B{y}_k (M)\Big]\label{eq:Y}
\end{equation}
The filter $\B{g}_k$ is then selected to
$$
\min_{\B{g}_k}\qquad \lVert\B{b}_k-\B{g}_k^{\He}\B{Y}_k\rVert^2 \label{eq:LSM}
$$
which gives
\begin{equation}
\B{g}_k = (\B{Y}_k\B{Y}_k^{\He})^{-1}\B{Y}_k\B{b}_k^{\He}.\label{eq:g_k}
\end{equation}



This is referred to as \emph{forward training}.
The beamformers $\B{v}_1 , \cdots , \B{v}_K$ are similarly
updated via \emph{backward training} exploiting channel reciprocity.
Specifically, the reverse channel from receiver $k$
to transmitter $j$ is
$\overleftarrow{\B{H}}_{jk} = \B{H}_{kj}^{\Tr}$.
Fixing the set of (original) receive filters $\{\B{g}_k\}$,
receiver $k$ then applies $\B{g}_k^*$ as the beamformer, and
all receivers synchronously transmit training sequences
in the reverse direction. Let $\overleftarrow{\B{b}}_k$
denote the training sequence from receiver $k$.
Then the observed signal at transmitter $k$ is given by
\begin{equation}
\overleftarrow{\B{Y}}_k =
\B{H}_{kk}^{(n)\Tr}\B{g}_k^*\overleftarrow{\B{b}}_k +
\sum_{j\neq k}\B{H}_{jk}^{(n)\Tr}\B{g}_j^*\overleftarrow{\B{b}}_j + \overleftarrow{\B{N}}_k
\end{equation}
where $\overleftarrow{\B{N}}_k =
\Big[\overleftarrow{\B{n}}_k^1,\cdots,\overleftarrow{\B{n}}_k^M\Big]$
is the vector of $M$ independent noise samples.

Note that $\overleftarrow{\B{Y}}_k^*$ corresponds to the
reverse signal used to compute the SINR in the Max-SINR algorithm,
where the transmitted symbol $x_k=\overleftarrow{\B{b}}_k^*$.
Hence replacing the corresponding MSE by the LS cost function,
we wish to select $\B{v}_k$ to
\begin{equation}
\min_{\B{v}_k}\qquad \Vert\overleftarrow{\B{b}}_k^* -
\B{v}_k^{\He} \overleftarrow{\B{Y}}_k^* \rVert^2 =
\lVert\overleftarrow{\B{b}}_k-\B{v}_k^{\Tr}
\overleftarrow{\B{Y}}_k\rVert^2 \label{eq:backward_LSM}
\end{equation}
giving the solution
\begin{equation}
\B{v}_k = \Big((\overleftarrow{\B{Y}}_k\overleftarrow{\B{Y}}_k^{\He})^{-1}
\overleftarrow{\B{Y}}_k\overleftarrow{\B{b}}_k^{\He}\Big)^*.\label{eq:v_k}
\end{equation}
We must normalize the beamformer/receive filter after
each update to satisfy the power constraint.
This does not change the SINR of the corresponding reverse/forward link.
However, it does influence the results in subsequent updates.
(This normalization is also included in the Max-SINR algorithm
and has been empirically observed to improve performance
relative to unnormalized updates.)


The {\em bi-directional} LS algorithm therefore consists of
the following steps:
\begin{enumerate}
\item \emph{Backward training}:
The receivers synchronously transmit $M$ training symbols given
by the backward training matrix $\overleftarrow{\B{B}}$. Receiver $k$ uses
the current estimate $\B{g}_k^*$ as the beamformer, and each transmitter $k$
updates the beamformer $\B{v}_k$ according to (\ref{eq:v_k}).

\item \emph{Forward training}:
The transmitters synchronously transmit $M$ training symbols
given by the training matrix $\B{B}$, and each receiver $k$
updates the filter $\B{g}_k$ according to (\ref{eq:g_k})
with a normalization to satisfy a power constraint.



\item Iterate the preceding steps up to a maximum number
of iterations.
\item Transmit data in the forward direction.

\end{enumerate}

The algorithm repeats each block. The training sequences
must be linearly independent across transmitters/receivers
in order to distinguish all sources, and ideally should
have low cross-correlation to improve the estimation accuracy.
As the training length $M$ becomes large,
the solution given by \eqref{eq:g_k} and \eqref{eq:v_k} approaches
the corresponding Minimum MSE solution, or equivalently,
the update in the Max-SINR algorithm.
Hence by running sufficiently many forward-backward cycles
within each block, each with sufficiently long training sequences,
the performance should approach that of
the Max-SINR algorithm.

For a fixed amount of training data there is generally
an optimal number of forward-backward iterations.
With too few iterations
the transmit beam and receiver filter do not converge
to the appropriate fixed point, whereas with too many iterations
each segment contains insufficient training symbols to obtain
accurate filter estimates. This is illustrated in the next
section. For high SNRs, the trade-off generally
favors more iterations since the number of iterations needed
to achieve the optimal fixed point increases with SNR.

When consecutive blocks
are highly correlated, i.e., $\alpha$ in \eqref{eq:blockfading}
is close to one, and with a fixed set of users,
the optimal number of iterations per block
is close to one in steady-state (i.e., once the beams and receivers
start to track the channel variations).
For this scenario the numerical results in
the next section assume that each block starts with forward training
using beams estimated from the preceding block.
Here we do not account
for the possibility that the receivers continue to train during
the data phase. If the receivers are able to track the channel
variations during the block, then further improvements are possible
by starting each block with backward training using the optimized
(updated) receivers as beams. (This also reduces the number
of switches between forward and backward training.)

With new users (or channels) the algorithm can be initialized
with either the forward or backward training phase.
Initializing with the backward phase may be best if the receivers
have {\em a priori} CSI (e.g., from previous transmissions).
Otherwise, the transmitters may initialize by transmitting pilots
through random beams.

As we noted previously, our bi-direction LS algorithm differs from other schemes for two-way channel estimation as in \cite{Mar06,GomPapSun08,OsaMurYamYos09,SteSab08,WitTayWar08,ZhoLamSad10} in that we directly estimate the filter coefficients as opposed to estimating the CSI needed to compute those coefficients.
The main advantages of direct filter estimation
are that it automatically accounts for varying interference levels
and filter estimation error. That is, pilots
from distant transmitters/receivers have little effect
on the filter estimate, so are automatically ignored.
In contrast, channel estimation schemes must
determine what CSI needs to be estimated.\footnote{This is
straightforward in a single-cell context
where CSI for all users in the cell is needed. However, deciding
on what CSI is important becomes an issue with multi-cell cooperation.}
Obtaining accurate CSI via two-way training may also
preclude all users from training
simultaneously, which extends the training period.
Finally, the filter estimation criterion
(e.g., least squares) provides the best filter estimate
at the transmit/receive side given the current set of
filters/beams at the opposite side. In contrast,
filter estimates must be modified to account
for inaccurate CSI \cite{GomPapSun08}.
The main disadvantage of the bi-directional filter estimation
scheme proposed here, relative to channel estimation, is
that it takes multiple iterations to converge.

\subsection{Short Coherence Blocks: Recursive Block Least Squares}
For a given coherence block length $L$ there is an optimal amount of
training per block; more training gives better filter estimates,
but takes away symbols for data transmission.
As the length of the coherence block decreases, the optimal training
length decreases. One way to effectively increase the amount of
training for small $L$ is to include training data from previous blocks.
Because the channels, beams, and filters are assumed to vary
over successive blocks, it is also important to discount
the data from past blocks when computing the current estimates.
One possibility is to modify the least squares cost function
by including exponentially weighted data from previous blocks, namely,
\begin{equation}
e_k^{(n)} = \sum_{l=1}^n\lambda^{(n-l)}
\bigg(\sum_{i=1}^M\lvert b_k^{(l)} (i)-
\B{g}_k^{\He}\B{y}_k^{(l)} (i)\rvert^2\bigg)
\end{equation}
where $n$ is the current block index,
$b_k^{(l)}(i)$ and $\B{y}_k^{(l)}(i)$ are, respectively, the $i$-th training symbol
and the corresponding received signal vector for
user $k$ in block $l$, the summation in the parentheses
is the sum error square for the $l$-th block,
and $\lambda\in(0,1]$ is the exponential weighting factor.
Roughly speaking, the memory of the algorithm spans
$1/(1-\lambda)$ coherence blocks.
Taking $\lambda=1$ corresponds to infinite memory.

We also add a regularization term $\delta\lambda^n\lVert\B{g}_k\rVert$,
where $\delta$ is a small positive constant. This helps
to stabilize the solution when $n$ is small, since
the amount of training may be insufficient to estimate the filters.
Similar to the LS algorithm in the preceding subsection,
for the receive filter update in the original direction,
given the training sequence $\B{b}_k^{(l)}$ and the
received signals $\B{Y}_k^{(l)}$ for block $l=1,\cdots,n$,
the receive filter of user $k$ is updated by solving
\begin{equation}
\min_{\B{g}_k} \quad \sum_{l=1}^n
\lambda^{n-l}\lVert\B{b}_k^{(l)}-\B{g}_k^{\He}\B{Y}_k^{(l)}\rVert^2+
\delta\lambda^n\lVert\B{g}_k\rVert^2. \label{eq:RLSM}
\end{equation}
The solution to this minimization problem can be computed
for each block $n$.
However, it is not necessary to store all of the past data
to update the solution. A block recursive algorithm for
updating $\B{g}_k$ is
shown in Table \ref{table1}, and consists of
updating the state variables $\B{P}_k^{(n)}$ ($N_R \times N_R$ matrix),
and calculating $\B{K}_k^{(n)}$ ($N_R \times M$ matrix)
at each block using the current training data.

\begin{table}[!h]\normalsize
\centering
\caption{Block Recursive LS Algorithm}\label{table1}
\begin{tabular}{l}
\hline
Initialization\\
\hline
$\B{g}_k^{(0)} = 0$\\
$\B{P}_k^{(0)} = \delta^{-1}\B{I}_{N_R\times N_R}$\\
\hline
For each block $n$, compute\\
\hline
$\B{K}_k^{(n)}=\lambda^{-1}\B{P}_k^{(n-1)}\B{Y}_k^{(n)}\Big(\B{I}_{M\times M}+$\\
$\qquad\qquad\lambda^{-1}\B{Y}_k^{(n)\He}\B{P}_k^{(n-1)}\B{Y}_k^{(n)}\Big)^{-1}$\\
$\B{g}_k^{(n)}=\B{g}_k^{(n-1)}-\B{K}_k^{(n)}(\B{b}_k^{(n)\He}-\B{Y}_k^{(n)\He}\B{g}_k^{(n-1)})$\\
$\B{P}_k^{(n)}=\lambda^{-1}\B{P}_k^{(n-1)}-\lambda^{-1}\B{K}_k^{(n)}\B{Y}_k^{(n)\He}\B{P}_k^{(n-1)}$\\
\hline
\end{tabular}
\end{table}

Similarly, in the backward direction, we update the beamformer
$\B{v}_k$ using the analogous exponentially weighted LS objective, i.e.,
\begin{equation}
\min_{\B{v}_k} \quad \sum_{l=1}^n \lambda^{n-l}\lVert\overleftarrow{\B{b}}_k^{(l)}-\B{v}_k^{\Tr}\overleftarrow{\B{Y}}_k^{(l)}\rVert^2+\delta\lambda^n\lVert\B{v}_k\rVert^2. \label{eq:backward_RLSM}
\end{equation}
The recursive method in Table \ref{table1} is again applicable
where each transmitter updates the matrix
$\B{Q}_k^{(n)}$ as the counterpart of $\B{P}_k^{(n)}$.
The \emph{Bi-Directional Recursive Least Squares (RLS)}
algorithm is then given by the following steps:
\begin{enumerate}
\item \emph{Initialization:}
The first training phase can be through
an arbitrary set of beamformers/receive filters, although
the updates in Table 1 are initialized by setting $\mathbf{g}_k^{(0)}$
to the all-zero vector,
$\B{P}_k^{(0)}=\delta^{-1}\B{I}_{N_R\times N_R}$, and
$\B{Q}_k^{(0)}= \delta^{-1}\B{I}_{N_T\times N_T}$.

\item \emph{Backward training:}
The receivers synchronously transmit $M$ training symbols
using the current normalized receivers as beams, i.e.,
receiver $k$ sends the training symbols
$\overleftarrow{\B{b}}_k^{(n)}$ to its associated
transmitter with the normalized version of the beam $\B{g}_k^{(n-1)*}$.
Each receiver $k$ then updates $\mathbf{v}_k^{(n)}$ according
to Table \ref{table1} substituting $\B{v}_k^*$ and $\B{Q}_k$
for $\B{g}_k$ and $\B{P}_k$.

\item \emph{Forward training:}
The transmitters synchronously transmit $M$ training symbols
using the current normalized beams, i.e.,
transmitter $k$ transmits the $M$ training
symbols $\B{b}_k^{(n)}$ using the normalized version of the
beam $\B{v}_k^{(n)}$.
Each receiver $k$ then updates $\mathbf{g}_k^{(n)}$,
$\B{K}_k^{(n)}$, and $\B{P}_k^{(n)}$
according to the algorithm in Table \ref{table1}.


\item The transmitters synchronously transmit data in the forward direction using the updated beams for the duration of the coherence block.
\end{enumerate}

In contrast with the (unweighted) LS algorithm, here we assume
that there is only one iteration per block. This is because
the training length is assumed to be relatively short, so
that multiple iterations would likely perform worse.

\section{Numerical Results}
In this section, we present numerical results for a network of three
users with $2\times2$ MIMO channels\footnote{As shown in
  \cite{CadambeJafarIT08}, interference alignment is achievable in
  this setting.}. In each simulation run, all
channel matrices (direct and cross) are
independently generated following the block-fading model in
\eqref{eq:blockfading} with unit variance and a given choice of $\alpha$.
White Gaussian additive noise is assumed with variance
$\sigma_0^2$ so that the SNR is then $1/\sigma_0^2$. All results shown
are averaged over multiple channel realizations. Next, we present
results for the bi-directional LS algorithm. We then consider the
bi-directional RLS algorithm. All training schemes presented in this section also include an additional receive filter update during the data transmission. Specifically, each receiver applies the current filter to estimate the transmitted binary symbols, and at the end of each block, updates the receive filter again to minimize the sum error square of those estimated symbols, similar to the forward training in the bi-directional LS algorithm. However, the sum rate is still evaluated at the beginning of the data transmission period for each block.

\subsection{Bi-directional Least Squares Algorithm}

\begin{figure}
\centering
\includegraphics[width=3.5in]{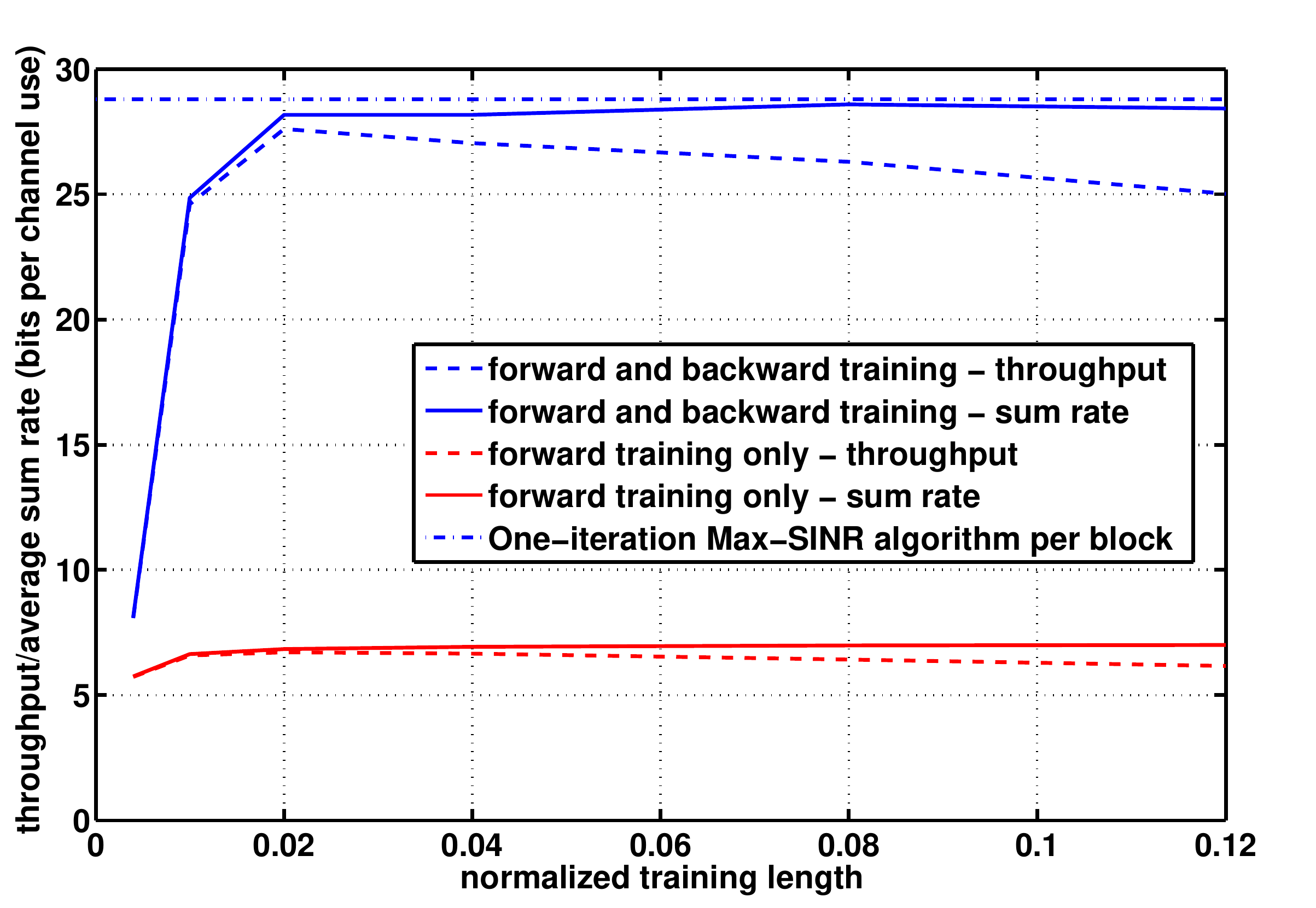}
\caption{Performance of the bi-directional LS algorithm and the
  forward training only scheme for constant channels.}\label{fig1}
\end{figure}

We first consider the bi-directional LS algorithm for
constant channels ($\alpha =1$) with a SNR of $30$~db.
In this case, as the training length
goes to infinity, the sum-rate achieved by the algorithm should
approach that achieved by the Max-SINR algorithm with perfect CSI.
This is illustrated in Figure~\ref{fig1}, which shows the sum-rate
achieved with bi-directional training and that achieved by the Max-SINR
algorithm as a function of the training length ($2M$)
normalized by $L$, which is 1000 symbols here. The figure also shows
the throughput achieved the bi-directional scheme, where
throughput is given by the rate per channel use after subtracting off the
overhead for training. After accounting for this overhead, it can be
seen that the optimal normalized training length is around 0.02.
A single iteration of bi-directional
training per block is applied in this example. For comparison,
we also show the sum-rate and throughput achieved by a ``forward
training only'' scheme in which
the initial beamformer of each user is fixed and only the receiver filters
are updated using by the same training method.  Clearly the two-way
training significantly improves over only one-way training.

\begin{figure}
\centering
\includegraphics[width=3.5in]{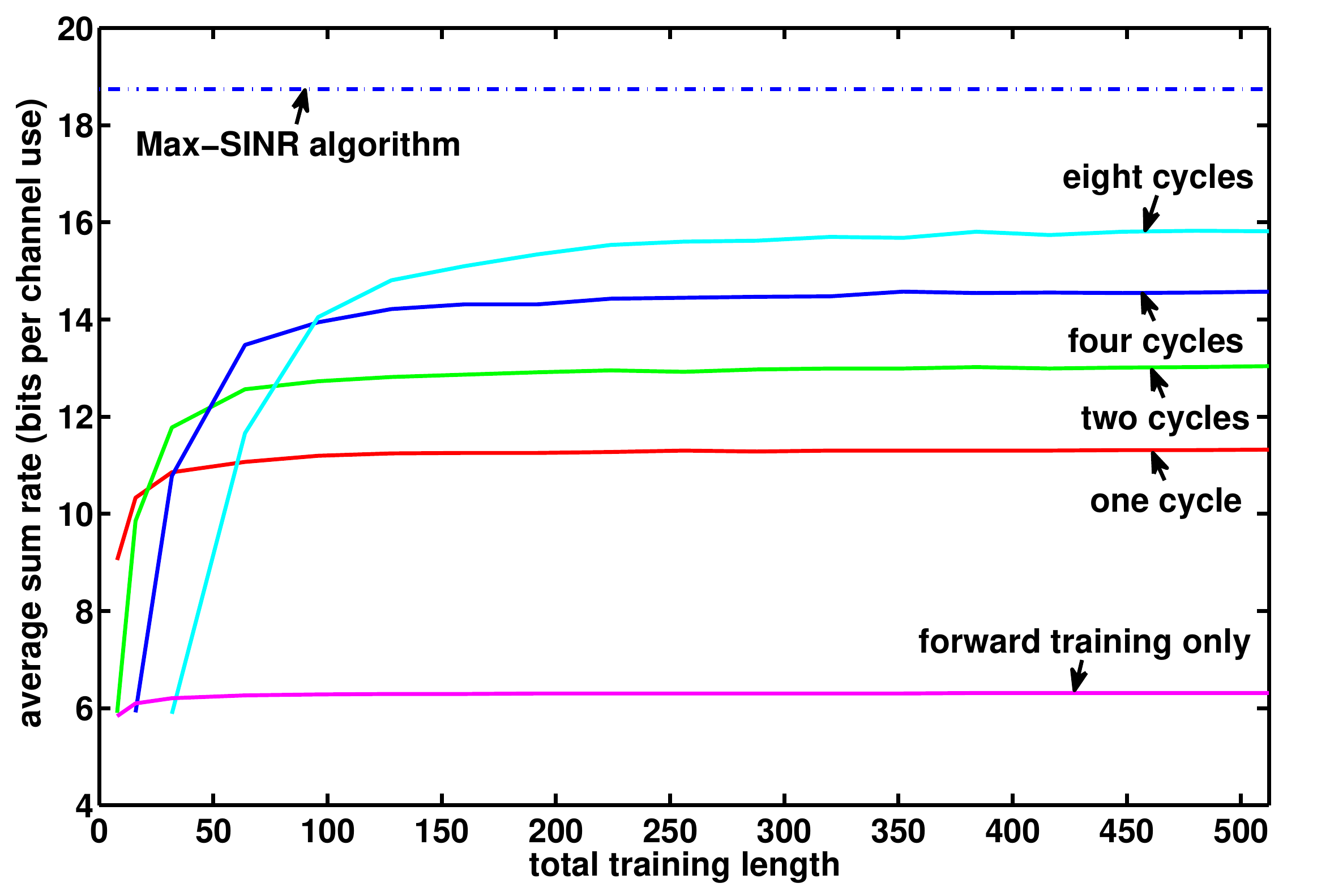}
\caption{Sum-rate versus training length with bi-directional training and \emph{i.i.d.} block
fading channels. Plots are shown for different numbers of forward-backward iterations (cycles).}\label{fig_multicycles}
\end{figure}

Next, to illustrate the effect of varying the number of iterations per
coherence block, we consider a network with \emph{i.i.d.} block fading channels
(i,e, $\alpha = 0$). Figure~\ref{fig_multicycles}
shows the sum-rate versus the total training length for a SNR of
$20$~dB. Each curve corresponds to a different number of iterations
(cycles) per block, where the total amount of training is evenly
divided among each iteration. For example, with 128 training symbols
and 4 bi-directional iterations, the training alternates between the
forward and reverse directions every 16 symbols. As in the previous
case, we also show the performance of the Max-SINR algorithm and that
of forward training only. Again for this case, the
bi-directional training scheme can provide a substantial benefit
relative to forward training only. However, the results also indicate
that significant training is needed to approach the  sum rate possible
with perfect channel knowledge. It can also be seen that given a fixed
training length there is an optimal number of bi-directional
iterations; with too few iterations the transmit beam and receiver
filter do not converge to the appropriate fixed point, whereas with
too many iterations each segment contains insufficient training
symbols to obtain accurate filter estimates. For high SNRs, the
tradeoff generally favors more iterations since the number of
iterations needed to achieve the optimal fixed point increases with
SNR.

\begin{figure}
\centering
\includegraphics[width=3.5in]{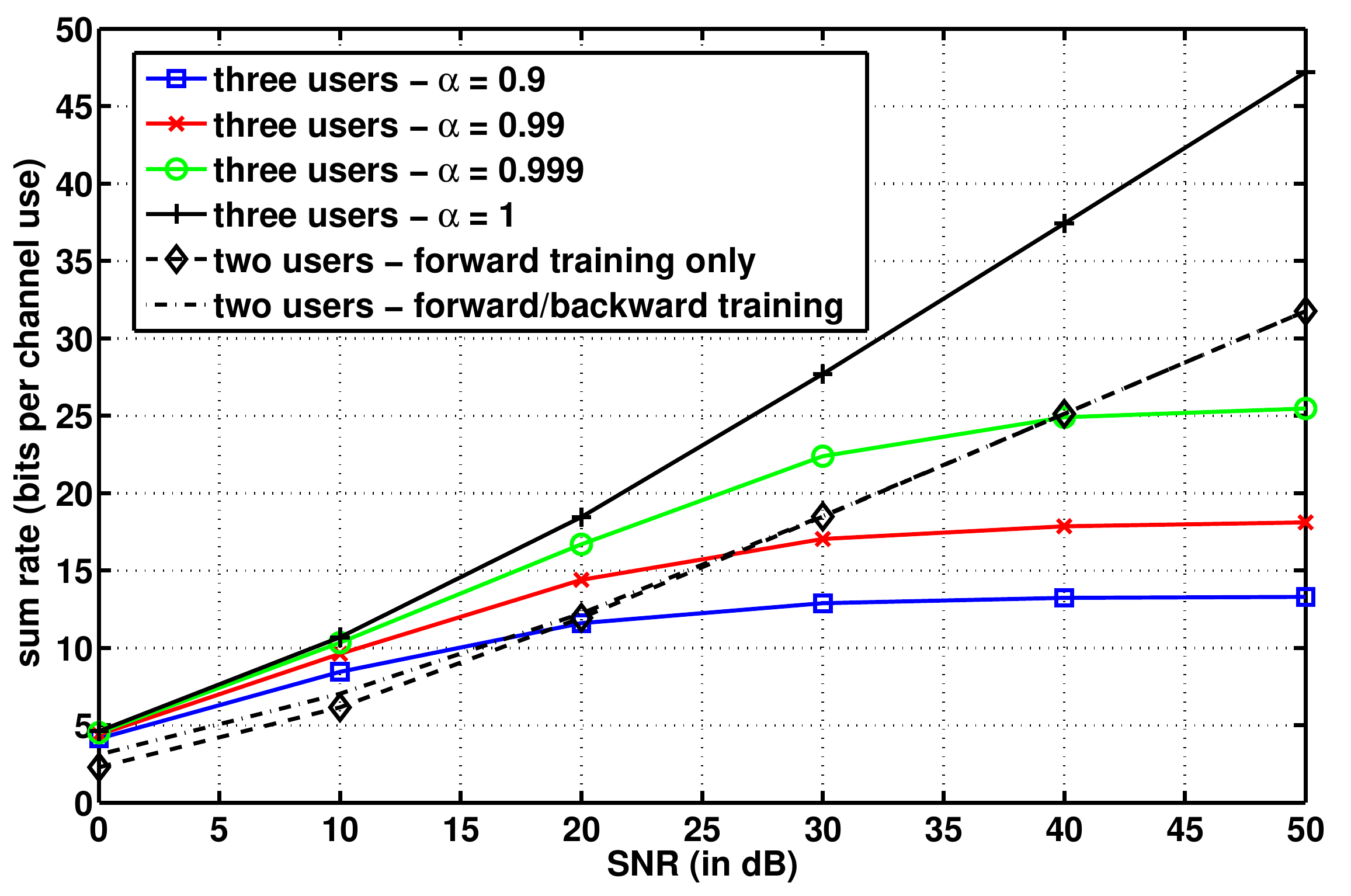}
\caption{Sum rate performance of the bi-directional LS algorithm
with one iteration and infinite training data per block
under different channel correlations.}\label{fig4}
\end{figure}

Next, to gain insight into the performance of the bi-directional LS
algorithm as a function of the SNR, we consider a limiting model in
which their is a single iteration per block and
the training data per iteration goes to infinity.
In this case,  each update will be equivalent to the corresponding update
in the Max-SINR algorithm. Fig.~\ref{fig4} shows the sum rate yielded
by this limiting scheme above versus the SNR for channels with
correlated fading corresponding to different choices of $\alpha$.
We also show the performance of two schemes in which only two of the
three users are allowed to transmit and the two users either update
their beams using bi-directional training or forward training only,
still assuming an infinite number of training bits per update.
First consider the scheme corresponding to three users with $\alpha
=1$. In this case the channel is not changing from one iteration to
the next and thus the algorithm is the same as the Max-SINR algorithm,
which appears to be achieving the optimal high-SNR slope
here. When the channel is not constant ($\alpha<1$), the
beamformers and receive filters are always
mismatched due to the channel fading; this mismatch eventually limits
the growth of the sum-rate for the three user schemes as the SNR increases.
The two schemes in which only two user transmit have nearly identical
performance and both appear to achieve the optimal high-SNR slope
for two users with rank 1 codebooks.\footnote{Fig.~\ref{fig4} only shows
  the curves for the two-user schemes
corresponding to the case of $\alpha = 0$; the performance for other
choices of $\alpha$ is very similar.}
This is because for a two-user $2\times 2$ MIMO system, it is much
easier to orthogonalize the two users and indeed this can be done by
only adapting the receive filters, while for a three-user system, both
the beamformers and receive filters need to be chosen to achieve alignment.
This suggests that for a high enough SNR in a time-varying channel,
it may be better to only allow two-users to transmit in this way.

\subsection{Bi-directional Recursive Least Squares Algorithm}

\begin{figure}
\centering
\includegraphics[width=3.5in]{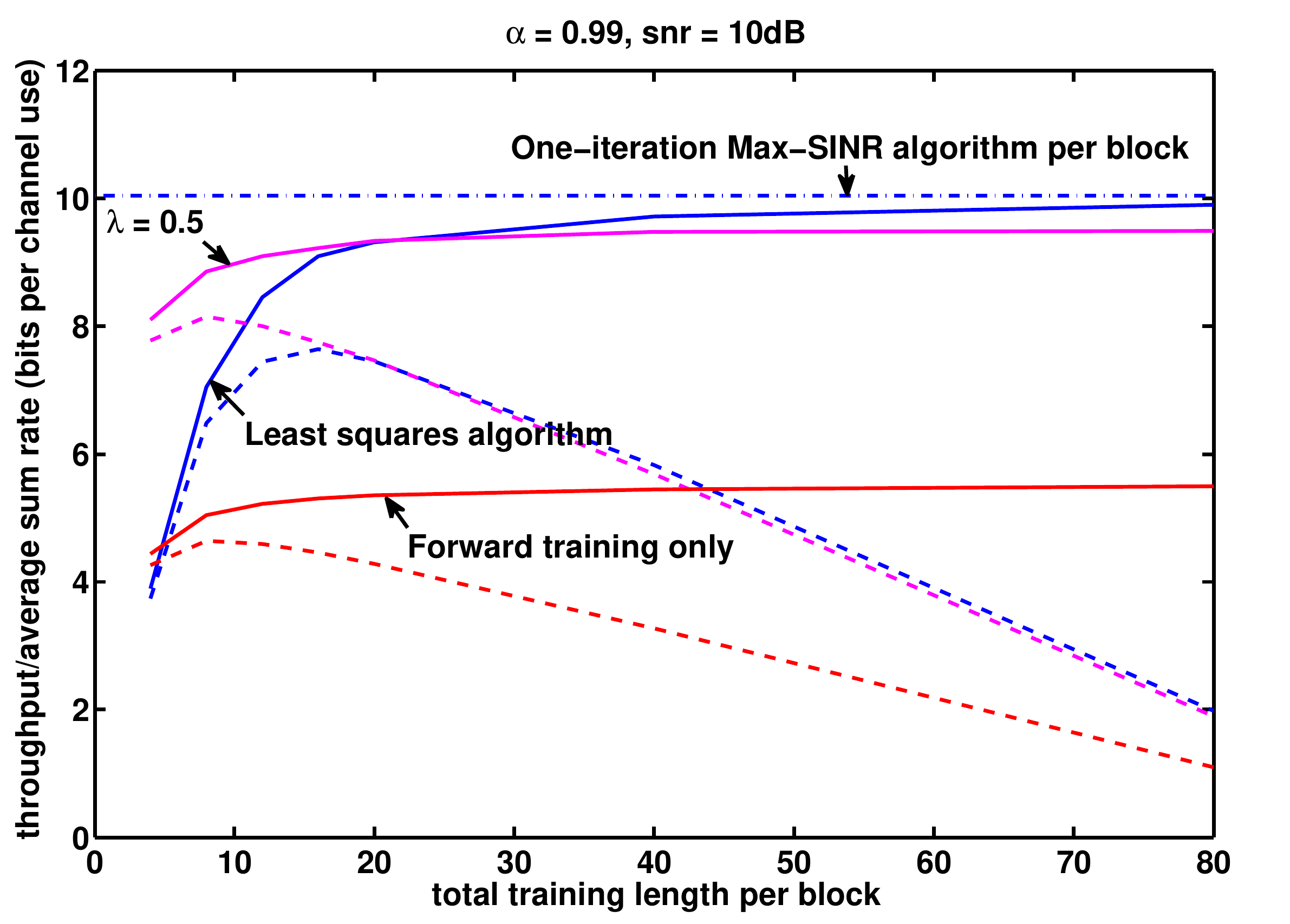}
\caption{Comparison between the LS algorithm and the RLS algorithm with $\lambda = 0.5$ ($\alpha = 0.99$).}\label{fig3}
\end{figure}

\begin{figure}
\centering
\includegraphics[width=3.5in]{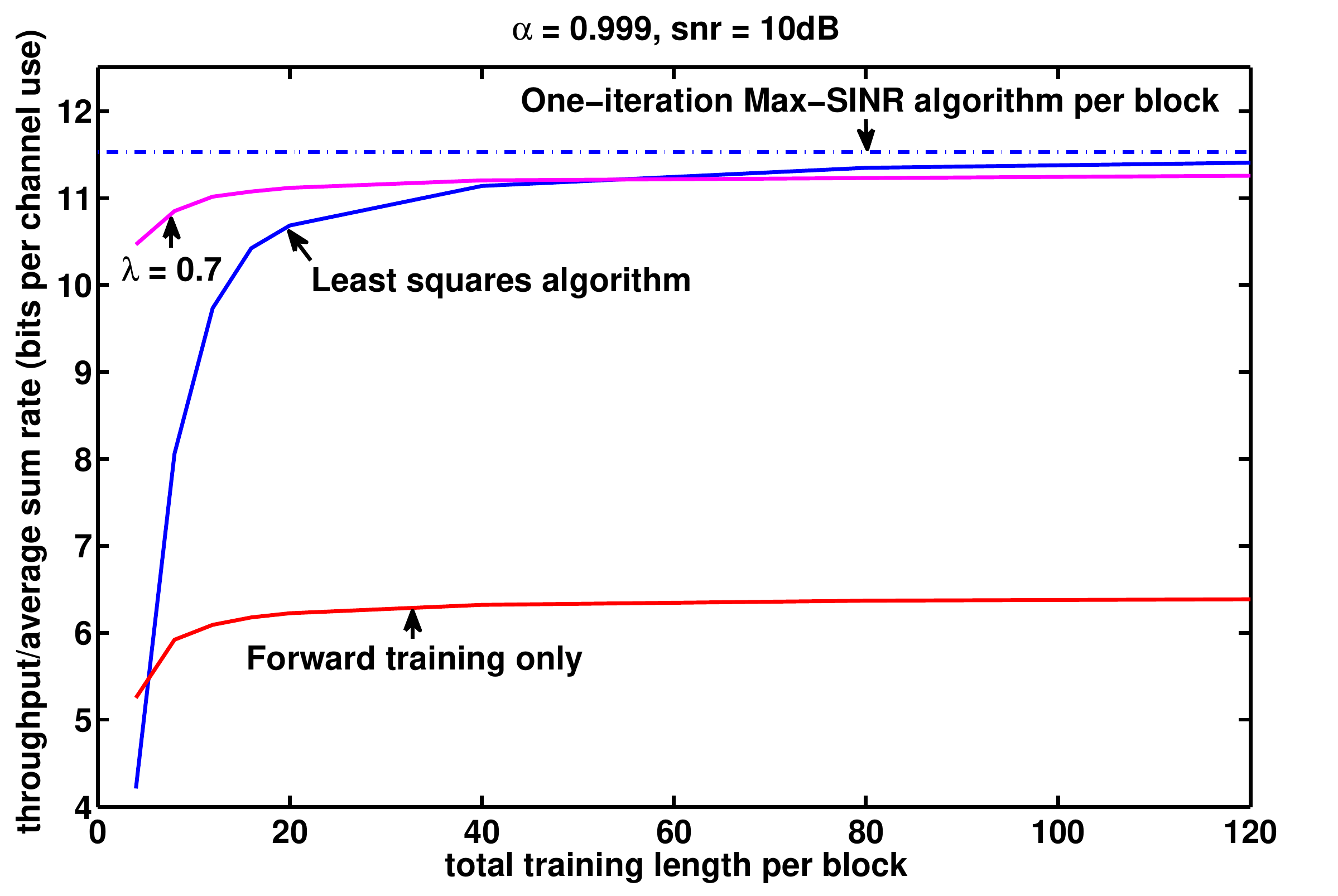}
\caption{Comparison between the LS algorithm and the RLS algorithm with $\lambda = 0.7$ ($\alpha = 0.999$).}\label{fig2}
\end{figure}

Now we turn to the performance of the
bi-directional RLS algorithm. Recall, that this algorithm was
motivated for cases where the number of channel uses per block is
small and there is significant correlation between blocks.
Figures~\ref{fig3} and \ref{fig2} show the performance of
the bi-directional RLS with different choices of $\lambda$ and
LS algorithms as a function of the training length in
a channels with correlated fading corresponding to $\alpha = 0.99$ and
$\alpha = 0.999$, respectively. Both the sum-rate (solid
lines) and the throughput (dashed lines) of each algorithm is shown as
well as for forward-training only. For all the algorithms,
a single iteration of training is used per block and the SNR
is 10db.  It can be seen that when the total training length is
limited the bi-directional RLS algorithm with an appropriate $\lambda$
gives a higher sum-rate than the LS algorithm, while if training bits are
sufficient the LS algorithm gives the high rate. The gains of the
bi-directional RLS algorithm are more significant
when $\alpha$ is closer to $1$, i.e., the channel is varying
more slowly.  When we consider the
throughput accounting for training overhead, the gains of the
bi-directional RLS algorithm diminish, and for $\alpha =0.99$ become
insignificant for most ranges of training. Of course this comparison
depends on the block-length $L$, which here we assume is given by
$L=\frac{1}{1-\alpha}$.
In other simulations, we have also observed that the performance
benefits of the RLS algorithm are greater at lower SNRs, i.e.~when
estimation becomes more difficult.

\section{Conclusions}
We have presented a distributed algorithm for iteratively adapting
beamformers and receive filters in MIMO interference networks without
CSI. The algorithm is based on using bi-directional training in a
synchronous TDD system. This
algorithm approaches the performance of the Max-SINR
algorithm with full CSI as the amount of training and the
number of training cycles increase. A recursive modification of the
algorithm that uses exponentially weighted data from previous blocks
was also given and shown to offer better performance when the channels
are highly correlated and the training length very small. Here we
mainly demonstrated the performance of these algorithms via simulations;
analyzing the performance is an interesting direction for future work.


\begin{thebibliography}{11}

\bibitem{JafarGlobecom08}
K. S. Gomadam, V. R. Cadambe, and S. A. Jafar, ``Approaching the Capacity of Wireless Networks through Distributed Interference Alignment,'' in \emph{Proc. of IEEE GLOBECOM}, Nov./Dec. 2008.

\bibitem{Asilomar09}
D. A. Schmidt, C. Shi, R. A. Berry, M. L. Honig, and W. Utschick, ``Minimum Mean Squared Error Interference Alignment,'' in \emph{Proc. of Asilomar Conference on Signals, Systems, and Computers}, Nov., 2009.

\bibitem{ICC09}
C. Shi, D. A. Schmidt, R. A. Berry, M. L. Honig, and W. Utschick, ``Distributed Interference Pricing for the MIMO Interference Channel,'' in \emph{Proc. IEEE International Conference on Communications}, June 2009.

\bibitem{CadambeJafarIT08}
V. R. Cadambe and S. A. Jafar, ``Interference Alignment and the Degrees of Freedom for the K User Interference Channel,'' \emph{IEEE Transactions on Information Theory}, vol.~54, no.~8, pp. 3425-3441, Aug. 2008.

\bibitem{PetersHeath09}
S. W. Peters and R. W. Heath, Jr., ``Cooperative Algorithms for MIMO Interference Channels,'' submitted to \emph{IEEE Transactions on Vehicular Technology}, Dec. 2009.

\bibitem{Mar06}
T. L. Marzetta, ``How much training is required for multiuser MIMO?''
In {\it Proc. Asilomar Conf. on Signals, Systems, and Computers}, pp.~359-363, Pacific Beach, Ca., 2006.

\bibitem{ZhoLamSad10}
X. Zhou, T. A. Lamahewa, P. Sadeghi, and S. Durrani, ``Two-Way Training: Optimal Power Allocation for
Pilot and Data Transmission,'' \emph{IEEE Trans. on Wireless Communications}, Feb. 2010.

\bibitem{GomPapSun08}
K. S. Gomadam, H. C. Papadopoulos, and C.-E. W. Sundberg, ``Techniques for Multi-user MIMO with Two-way Training,'' in \emph{Proc. IEEE International Conference on Communications}, May, 2008.

\bibitem{OsaMurYamYos09}
R. Osawa, H. Murata, K. Yamamoto, and S. Yoshida, ``Performance of Two-Way Channel Estimation Technique for Multi-User Distributed Antenna Systems with Spatial Precoding,'' in \emph{Proc. Vehicular Technology Conference Fall (VTC 2009-Fall)}, Sept. 2009.

\bibitem{SteSab08}
C. Steger and A. Sabharwal, ``Single-input Two-way SIMO Channel: Diversity-multiplexing Tradeoff with Two-way Training,'' \emph{IEEE Transactions on Wireless Communications}, Dec. 2008.

\bibitem{WitTayWar08}
L. P. Withers, R. M. Taylor, and D. M. Warme, ``Echo-MIMO: A Two-Way Channel Training Method for Matched Cooperative Beamforming,'' \emph{IEEE Transactions on Signal Processing}. Sept. 2008.





%
%
%
%
%
%
%
%
%
%
%

\end{thebibliography}
\end{document}